\documentclass[aps,prb,twocolumn,floatfix,superscriptaddress]{revtex4}
\usepackage{times}
\usepackage{graphicx}
\usepackage{epsfig}
\usepackage{amsmath}
\usepackage{amssymb}
\usepackage[version=3]{mhchem}

\begin{document}

\title{Hybrid Local-Order Mechanism for Inversion Symmetry Breaking}

\author{Emma H. Wolpert}
\affiliation{Department of Chemistry, University of Oxford, Inorganic Chemistry Laboratory, South Parks Road, Oxford OX1 3QR, U.K.}

\author{Alistair R. Overy}
\affiliation{Department of Chemistry, University of Oxford, Inorganic Chemistry Laboratory, South Parks Road, Oxford OX1 3QR, U.K.}
\affiliation{Diamond Light Source, Chilton, Oxfordshire, OX11 0DE, U.K.}

\author{Peter M. M. Thygesen}
\affiliation{Department of Chemistry, University of Oxford, Inorganic Chemistry Laboratory, South Parks Road, Oxford OX1 3QR, U.K.}

\author{Arkadiy Simonov}
\affiliation{Department of Chemistry, University of Oxford, Inorganic Chemistry Laboratory, South Parks Road, Oxford OX1 3QR, U.K.}

\author{Mark S. Senn}
\affiliation{Department of Chemistry, University of Warwick, Gibbet Hill, Coventry CV4 7AL, U.K.}

\author{Andrew L. Goodwin$^\ast$}
\affiliation{Department of Chemistry, University of Oxford, Inorganic Chemistry Laboratory, South Parks Road, Oxford OX1 3QR, U.K.}

\date{\today}
\begin{abstract}
Using classical Monte Carlo simulations, we study a simple statistical mechanical model of relevance to the emergence of polarisation from local displacements on the square and cubic lattices. Our model contains two key ingredients: a Kitaev-like orientation-dependent interaction between nearest neighbours, and a steric term that acts between next-nearest neighbours. Taken by themselves, each of these two ingredients is incapable of driving long-range symmetry breaking, despite the presence of a broad feature in the corresponding heat capacity functions. Instead each component results in a ``hidden'' transition on cooling to a manifold of degenerate states; the two manifolds are different in the sense that they reflect distinct types of local order. Remarkably, their intersection---\emph{i.e.} the ground state when both interaction terms are included in the Hamiltonian---supports a spontaneous polarisation. In this way, our study demonstrates how local ordering mechanisms might be combined to break global inversion symmetry in a manner conceptually similar to that operating in the ``hybrid'' improper ferroelectrics. We discuss the relevance of our analysis to the emergence of spontaneous polarisation in well-studied ferroelectrics such as BaTiO$_3$ and KNbO$_3$.
 \end{abstract}


\maketitle

\section{Introduction}

Central to the study of ferroelectric materials is an understanding of the collective mechanisms responsible for inversion-symmetry breaking in solids.\cite{Jona_1962,Scott_1998} For proper ferroelectrics (\emph{e.g.}\ the long-studied PbTiO$_3$),\cite{Shirane_1970} the key collective behaviour is a single zone-centre polar phonon that softens on cooling the paraelectric parent below the Curie temperature, $T_{\textrm C}$. A conceptually similar picture emerges in improper ferroelectrics such as Gd$_2$(MoO$_4$)$_3$ and YMnO$_3$, where the polar phonon instability is driven by a non-polar distortion that actually acts as the primary order parameter.\cite{Levanyuk_1974,vanAken_2004,Fennie_2005}

An important recent development in the field has been the \emph{hybrid} improper ferroelectric (HIF) mechanism relevant to \emph{e.g.}\ Ca$_3$Mn$_2$O$_7$.\cite{Bousquet_2008,Benedek_2011,Oh_2015,Xu_2015} The remarkable feature of this mechanism is that spontaneous polarisation emerges \emph{via} trilinear coupling to two non-polar distortion modes. The important implication in the particular context of multiferroics is that polarisation--magnetisation coupling might be targetted through judicious combinations of non-polar distortions (\emph{e.g.}\ tilts or cation order) rather than attempting to reconcile the inherently antagonistic requirements for magnetic order (partially-filled $d$ orbitals) and polar instabilities (closed-shell $d$-electron configurations).\cite{Hill_2000} Such ``tilt-engineering'' approaches have now been exploited to remarkable effect in the design of (Ca$_y$Sr$_{1-y}$)$_{1.15}$Tb$_{1.85}$Fe$_2$O$_7$, which shows spontaneous polarisation and magnetisation at room temperature.\cite{Pitcher_2015}

\begin{figure}
\begin{center}
\includegraphics{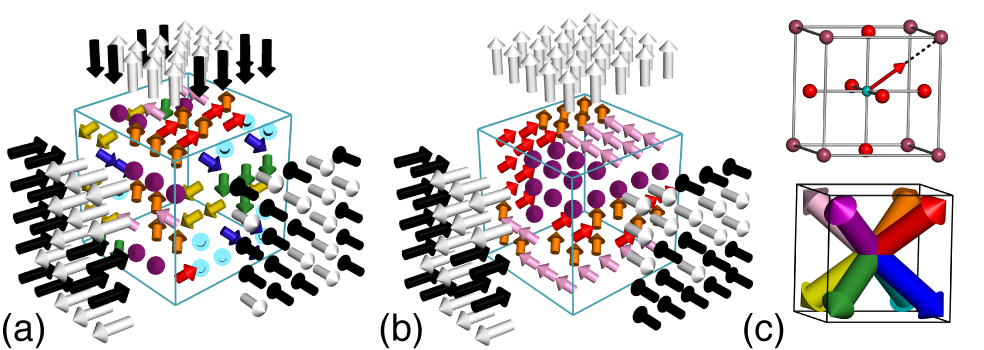}
\end{center}
\caption{\label{fig1} Schematic representations of local and collective polarisation in the (a) cubic and (b) tetragonal phases of BaTiO$_3$ or KNbO$_3$.\cite{Senn_2016} The dipole moments within individual perovskite cells are shown as coloured arrows. The collective polarisation for a given column (parallel to any one of the three crystal axes) is shown as a black or white arrow. In both phases there is strong local order (common polarisation projection along a given axis) but macroscopic inversion symmetry is broken only in the tetragonal phase. (c) Local polarisation is driven by a second-order Jahn-Teller distortion of (\emph{e.g.}) the TiO$_6$ coordination environment in which the Ti atom (teal sphere) displaces towards a single corner of the corresponding unit cell (red arrow). The eight possible displacement vectors $\langle111\rangle$ are shown in various colours.}
\end{figure}

In all these mechanisms, the collective distortions responsible for inversion-symmetry breaking---be they polar instabilities, octahedral tilts, cation ordering, or vacancy ordering processes---are each associated with individual irreducible representations of the parent crystal symmetry. In other words, polarisation emerges as a result of the relation
\begin{equation}\label{irrep}
\Gamma^-\subseteq\oplus_i\Phi_i^\pm,
\end{equation}
where $\Gamma^-$ is a (generic) polar distortion and the collective distortion space spanned by the direct sum is generated by a small and finite number of unstable distortion modes (possibly non-polar) with irredicuble representations $\Phi_i^\pm$. For proper ferroelectrics Eq.~\eqref{irrep} reduces to $\Gamma^-=\Phi^-$. While this is the conventional picture for systems such as BaTiO$_3$ and KNbO$_3$,\cite{Cochran_1960} recent total scattering measurements\cite{Senn_2016} have supported the hypothesis of Refs.~\citenum{Comes_1968,Comes_1970} that the paraelectric/ferroelectric transition may be strongly order/disorder in nature. In this alternate picture the ferroelectric transition involves a complex reorganisation of cation positions that cannot properly be described in the terms of Eq.~\eqref{irrep}. This is because both the paraelectric (cubic) and polar (tetragonal) phases support different types of strongly-correlated disorder that are not related to one another by activation of a small and finite set of distortion modes [Fig.~\ref{fig1}]. Instead the two states are related by the presence or absence of a particular type of \emph{local} order.\cite{Comes_1970,Senn_2016}

In this study, we explore the concept of establishing HIF mechanisms based on coupling of polarisation to distortions arising from local order. Our approach is to develop a simple microscopic model containing just two ingredients, each of which drives its own form of local ordering to a non-polar disordered phase with a manifold of degenerate ground states. These two ingredients are our local analogues of \emph{e.g.}\ the tilt and rotation distortions in the HIF mechanism of Ca$_3$Mn$_2$O$_7$.\cite{Benedek_2011} Whereas the individual HIF distortions are localised in $\mathbf k$-space and collective in direct space, the two ingredients of our model drive order that is localised in $\mathbf r$-space and collective in reciprocal space. Using Monte Carlo (MC) simulations, we show that the full Hamiltonian has as its ground state a disordered, but polar, phase that is conceptually related to tetragonal BaTiO$_3$ or KNbO$_3$. Hence a collective polar distortion emerges from coupling to two non-polar local-order instabilities. The key implication of this result is that judicious control over local ordering phenomena might offer an alternative route to as-yet unexplored classes of ferroelectric materials.

Our paper is arranged as follows. We begin by introducing the general model at the heart of our study, relating its ingredients to simple physical interactions likely to be relevant to real materials. We describe our MC approach and proceed to establish the phase behaviour of the two individual components of our Hamiltonian, demonstrating the existence of local-structure transitions in each case. Our focus then shifts to the behaviour of the combined interaction model. We determine a general phase diagram and show explicitly the coupling of local order to macroscopic polarisation. For ease of interpretation, we focus in these studies on a two-dimensional representation of the Hamiltonian. Nevertheless, our paper concludes with a discussion of the (straightforward) extension to three dimensions, which allows us to draw comparisons to physical systems such as BaTiO$_3$ and KNbO$_3$.



\section{Results and Discussion}


\subsection{Interaction model}

\begin{figure}[b]
\begin{center}
\includegraphics{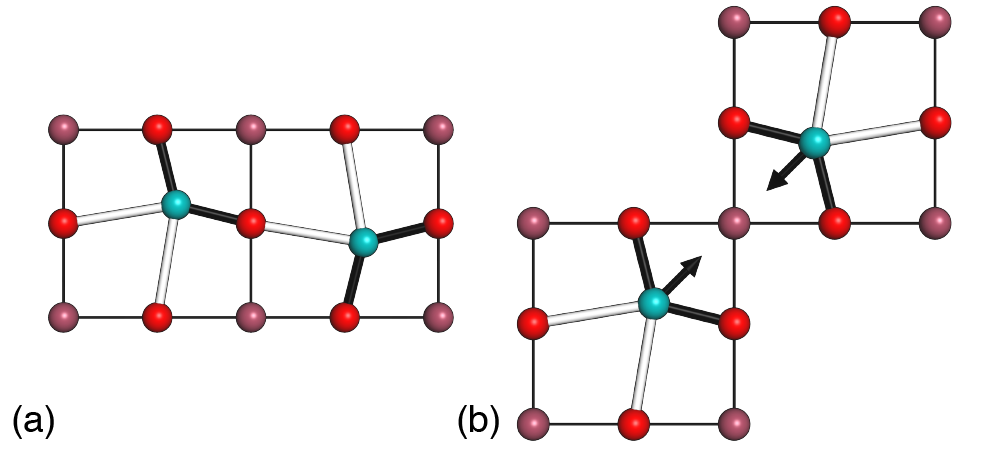}
\end{center}
\caption{\label{fig2} (a) A square ABO$_2$ lattice with alternating strong (black) and weak (white) B--O bonds. A-site cations are shown in brown, B-site in teal, and O in red. Note that each O atom is involved in exactly one strong and one weak B--O bond; this results in an alternating pattern of strong--weak--strong--$\ldots$ along each axis. (b) The steric interaction term of Eq.~\eqref{Hamil} penalises the displacement of two nearest-neighbour (diagonally-related) B-site cations towards one another---and hence towards a single common A-site cation.}
\end{figure}

Our 2D model concerns a square array of interacting sites. Each site $\mathbf r$ may adopt one of four equivalent states $\mathbf e$, which we associate with constant-magnitude off-center displacements polarised along the square diagonals [Fig.~\ref{fig2}(a)]; hence $\mathbf e_{\mathbf r}\in\langle11\rangle$ in two dimensions. In a physical system, this local displacement may be associated with (by way of example) a second-order Jahn Teller (SOJT) distortion or chemical bonding asymmetry; we are considering the particular case where all local displacements are of equal magnitude and are strictly polarised along the cell diagonals. Note that a simple coupling $J\mathbf e_{\mathbf r}\cdot\mathbf e_{\mathbf r^\prime}$ between neighbouring sites $\mathbf r,\mathbf r^\prime$ would result in a polar instability for $J<0$, but would give non-polar antiferroelectric and disordered ground states for $J>0$ and $J=0$, respectively.

The full Hamiltonian of our model does not include this particular interaction; instead we have
\begin{equation}\label{Hamil}
\mathcal H=-J_1\sum_{\mathbf r}\left(e_{\mathbf r}^xe_{\mathbf r+\mathbf a}^x+e_{\mathbf r}^ye_{\mathbf r+\mathbf b}^y\right)+J_2\sum_{\mathbf r}\delta(\mathbf e_{\mathbf r}+\mathbf e_{\mathbf r+\mathbf e_\mathbf r}),
\end{equation}
with $J_1,J_2\geq0$, $\delta$ the Kronecker delta function, and $\mathbf a,\mathbf b$ the lattice vectors.

The first term in Eq.~\eqref{Hamil} is conceptually related to the Kitaev model\cite{Kitaev_2006} of relevance to RuCl$_3$ and Na$_2$IrO$_3$:\cite{Ran_2017,Choi_2012} in both cases the interaction is bond-dependent. It describes a discrete ``compass'' model\cite{Nussinov_2015,Mishra_2004} that separates the 2D lattice into a set of non-interacting 1D Ising chains. Nearest neighbours separated by $\pm\mathbf a$ interact \emph{via} the $x$ component of their states (\emph{i.e.}, $e^x_{\mathbf r},e^x_{\mathbf r^\prime}$); likewise those separated by $\pm\mathbf b$ interact \emph{via} the $y$ components $e^y_{\mathbf r},e^y_{\mathbf r^\prime}$. Since $J_1>0$, energy is minimised whenever neighbouring states displace in the same direction relative to the axis along which they are connected. This interaction---if perhaps obscure at first sight---has a sound chemical basis for systems with SOJT distortions. Taking the example of a square ABO$_2$ lattice, such distortions would involve mixing of empty B $d$ states with filled O $p$ states to give a neighbouring pair of strong B--O bonds for each B site; the requirement that each O atom is involved in exactly one strong B--O bond gives precisely the Kitaev-like interaction present in our model [Fig.~\ref{fig2}(a)]. 

The second term in Eq.~\eqref{Hamil} penalises next-nearest neighbour displacements that act in direct opposition. This is a crude representation of the steric interaction that would arise from correlated B-site displacements towards a common A-site cation [Fig.~\ref{fig2}(b)]. The same type of next-nearest neighbour displacement pattern is penalised by a dipolar interaction---which one might argue is a more physical ingredient in a Hamiltonian---but the corresponding ground state is ordered and hence unsuitable for the purposes of our study. We anticipate that screened dipolar interactions probably account for the interaction in some physical systems.

In this context, and for ease of discussion, we hereafter denote the two components of Eq.~\eqref{Hamil} as `Kitaev' and `steric' interactions. We proceed to study the phase behaviour these interactions drive---first in isolation, and then in tandem.

\subsection{Monte Carlo simulations}

Our MC simulations made use of a custom-written code based on that employed in Ref.~\citenum{Paddison_2015} and were carried out as follows. A starting configuration corresponding to a $30\times30$ supercell of the square unit cell was decorated randomly with states $\mathbf e_{\mathbf r}\in\{[1,1],[1,\bar1],[\bar1,1],[\bar1,\bar1]\}$. We made use of periodic boundary conditions and followed the standard Metropolis Monte Carlo algorithm;\cite{Metropolis_1953} moves involved random reassignment of the state of a randomly-selected site, and energies $E$ were calculated for a given set of $J_1,J_2$ parameters according to Eq.~\eqref{Hamil}. Simulations were started at high temperatures and slowly cooled. We ensured equilibration at each temperature step by discarding ten times as many moves as were required for the system to become uncorrelated from its initial state ($t_0$). We then performed a further $100t_0$ moves to calculate the specific heat from the fluctuation--dissipation relation
\begin{equation}
C=\frac{\langle E^2\rangle-\langle E\rangle^2}{k_{\textrm B}T^2},
\end{equation}
where $k_{\textrm B}$ is the Boltzmann constant. Results were averaged over 5 independent runs.


\subsection{Behaviour of limiting cases}

\begin{figure}
\begin{center}
\includegraphics{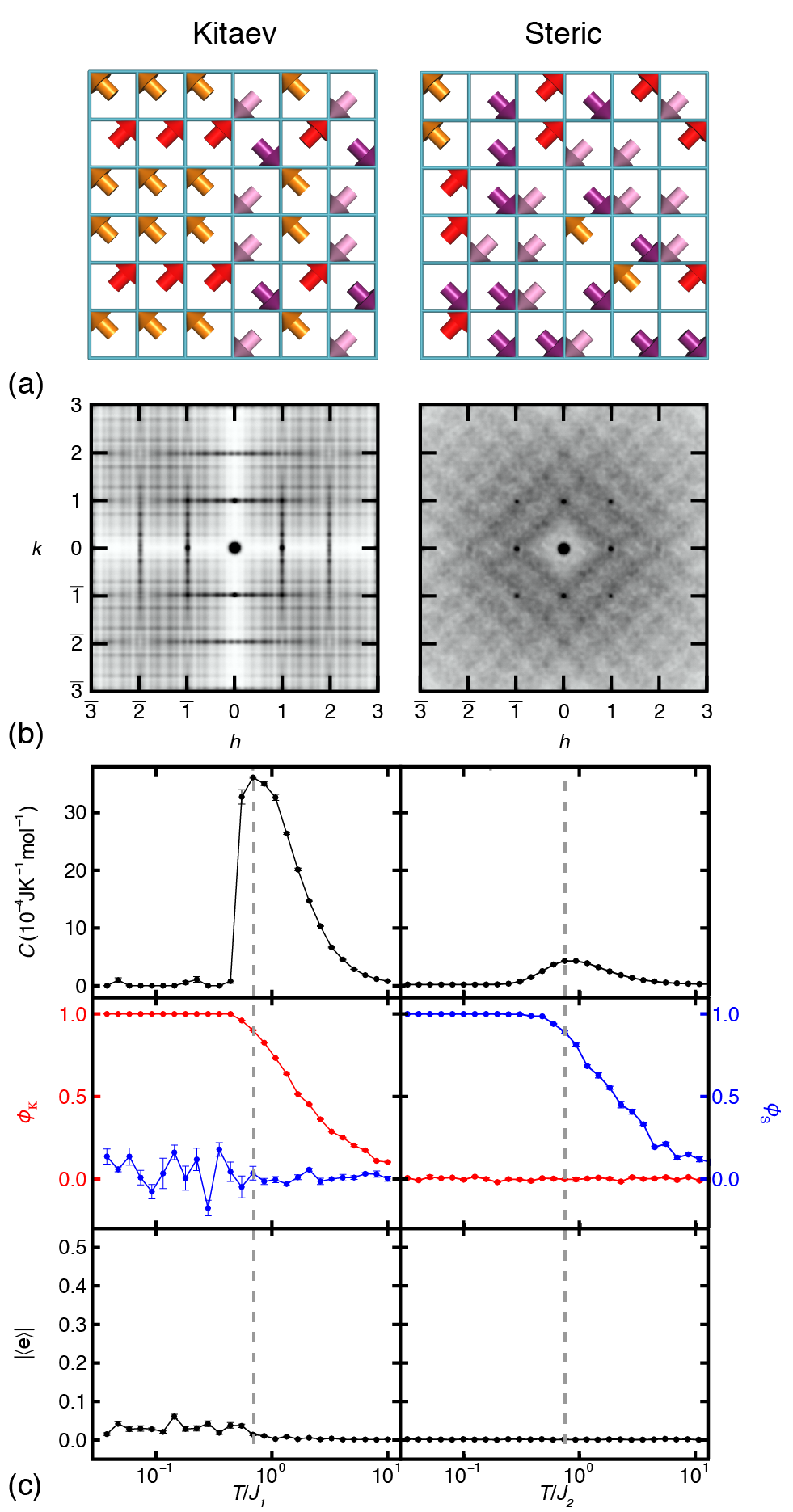}
\end{center}
\caption{\label{fig3} Microscopic phase behaviour in the single-term ``steric'' ($J_1=0$) and ``Kitaev'' ($J_2=0$) models described by Eq.~\eqref{Hamil}. (a) Fragments of the low-temperature displacement configurations are disordered in both cases, but contain different types of local order. The steric model contains no next-nearest neighbour displacements of the form depicted in Fig.~\ref{fig2}(b) and in the Kitaev model each row or column shares a common polarisation along the corresponding row or column axis. (b) The corresponding single-crystal scattering patterns are sensitive to two-particle correlation functions and reflect non-trivial disorder of different types in the two low-temperature configurations. The Bragg intensities (a measure of the single-particle correlations) are identical in the two cases. (c) Thermodynamic and ordering behaviour of the variable-temperature MC simulations. The top panels show the temperature dependence of the specific heat with the position of its maximum indicated by a vertical dashed line. The middle panels show the temperature dependence of the local order parameters $\phi_{\textrm K}$ (red) and $\phi_{\textrm s}$ (blue). Note the emergence of local-order of (only) one type at temperatures below the corresponding specific heat anomaly. The bottom panels show the macroscopic polarisation magnitude $|\langle\mathbf e\rangle|$, which is zero within error for all MC temperatures.}
\end{figure}

We begin by considering the phase behaviour of the pure Kitaev ($J_2=0$) and pure steric ($J_1=0$) interaction models. In both cases our MC simulations identified the existence of a single broad anomaly in the specific heat that we will come to show accompanies a meaningful change in local structure [Fig.~\ref{fig3}]. The maximum of this anomaly occurs at $T/J_i\simeq\frac{1}{2}$. In each case there is no long-range symmetry breaking at any finite temperature: this is is evident in the single-particle correlation function $\langle\mathbf e\rangle$, the magnitude of which vanishes within uncertainty for both models at all temperatures. This quantity is proportional to the bulk polarisation, which is therefore also zero for both models at all temperatures.

Despite the absence of long-range symmetry breaking in either transition, our thermodynamic calculations clearly show the loss of configurational entropy on cooling that we associate with local ordering processes. To illustrate this point we calculate two ``local-order parameters'' for each MC configuration:
\begin{eqnarray}
\phi_{\textrm K}&=&\frac{1}{4}\left[\langle(e^x_{\mathbf r}+e^x_{\mathbf r+\mathbf a})^2+(e^y_{\mathbf r}+e^y_{\mathbf r+\mathbf b})^2\rangle-4\right],\\
\phi_{\textrm s}&=&1-4\langle\delta(\mathbf e_{\mathbf r}+\mathbf e_{\mathbf r+\mathbf e_\mathbf r})\rangle,
\end{eqnarray}
with the average taken over all sites $\mathbf r$. By design, the value of $\phi_{\textrm K}$ is unity for any configuration in which all pairs of neighbouring states displace in the same direction relative to the axis along which they are connected, and zero for a random set of displacements $\mathbf e_{\mathbf r}$. Likewise a value $\phi_{\textrm s}=1$ implies the complete absence of any next-nearest neighbour displacements acting in direct opposition and $\phi_{\textrm s}=0$ the random case. The temperature dependence of these two parameters for the Kitaev and steric interaction models is shown in Fig.~\ref{fig3}(c); we observe a clear progression from an uncorrelated paraelectric state ($\phi\sim0$) at high temperatures to distinct locally-ordered paraelectric states at temperatures below the relevant specific heat anomalies.

The presence of different types of correlated disorder in the two limiting-case models at low temperatures is evident from the corresponding diffraction patterns\cite{Keen_2015} shown in Fig.~\ref{fig3}(b). The Bragg intensities for the two models are identical, which is necessarily the case given the absence of any variation in single-particle correlation functions across the specific heat anomalies. Instead the two families of low-temperature states differ only in terms of the pair (and higher-order) correlation functions. In this sense the local-order transitions we observe are ``hidden'',\cite{Okazaki_2011,Paddison_2015} and are conceptually related to the spin-ice transition in \emph{e.g.} Ho$_2$Ti$_2$O$_7$ and the superstructure transition of some inclusion compounds.\cite{Castelnovo_2008,Toudic_2008} Such transitions are not well described in terms of the conventional Landau paradigm, and can sometimes be viewed instead as a Higgs transition of an emergent gauge theory.\cite{Powell_2011}

Representative configurations themselves are illustrated in Fig.~\ref{fig3}(a); we note that the Kitaev phase is related to the ``S$_2$C'' procrystalline state as described in Ref.~\citenum{Overy_2016} and observed experimentally in perovskite oxynitrides.\cite{Yang_2011,Camp_2012} This state supports strong 1D correlations along the lattice axes---as is evident in the diffraction pattern \emph{via} the presence of continuous streaks of diffuse scattering perpendicular to the reciprocal lattice axes and is expected from the relationship to a decoupled 1D Ising model. Indeed the specific heat anomaly for this model [Fig.~\ref{fig3}(c)] is equivalent to that obtained for the 1D Ising case. In both cases there is no strict long-range 1D order at low temperatures; rather, regions of medium-range 1D order (\emph{i.e.}\ over many unit cells, with lengthscale inversely proportional to $T$) are separated by domain boundaries (Ising spin flips). With reference to our MC configurations, 1D order appears when the ordering lengthscale is commensurate with our simulation box size.

The low-temperature state of the steric interaction model is rather more disordered. This is evident in the diffraction pattern itself [Fig.~\ref{fig3}(b)], for which the diffuse scattering component is only weakly-structured throughout reciprocal space. Nevertheless the state is not random; the form of local order present involves the complete absence of next-nearest neighbour interactions of the type shown in Fig.~\ref{fig2}(b).

So, by themselves, each of the Kitaev and steric interaction terms in Eq.~\eqref{Hamil} drives a non-polar instability of the parent paraelectric state with respect to local ordering: these instabilities are localised in real-space (\emph{i.e.}\ strong local order) and collective in reciprocal space (\emph{i.e.}\ associated with large families of $\mathbf k$-points). In particular, the phase behaviour of both models cannot be described in terms of the activation of a small and finite set of collective (phonon-like) distortion modes. 

\subsection{Behaviour of intermediate case}

We now consider the effect of coupling Kitaev and steric interactions by studying the behaviour of our Hamiltonian for non-zero values of both $J_1$ and $J_2$. We define the parameters
\begin{eqnarray}
\theta&=&\tan^{-1}\left(\frac{J_2}{J_1}\right),\label{theta}\\
J&=&J_1+J_2,
\end{eqnarray}
such that the pure-Kitaev and pure-steric interaction models explored in the previous section correspond to the cases $\theta=0$ and $\pi/2$, respectively.

\begin{figure}[b]
\begin{center}
\includegraphics{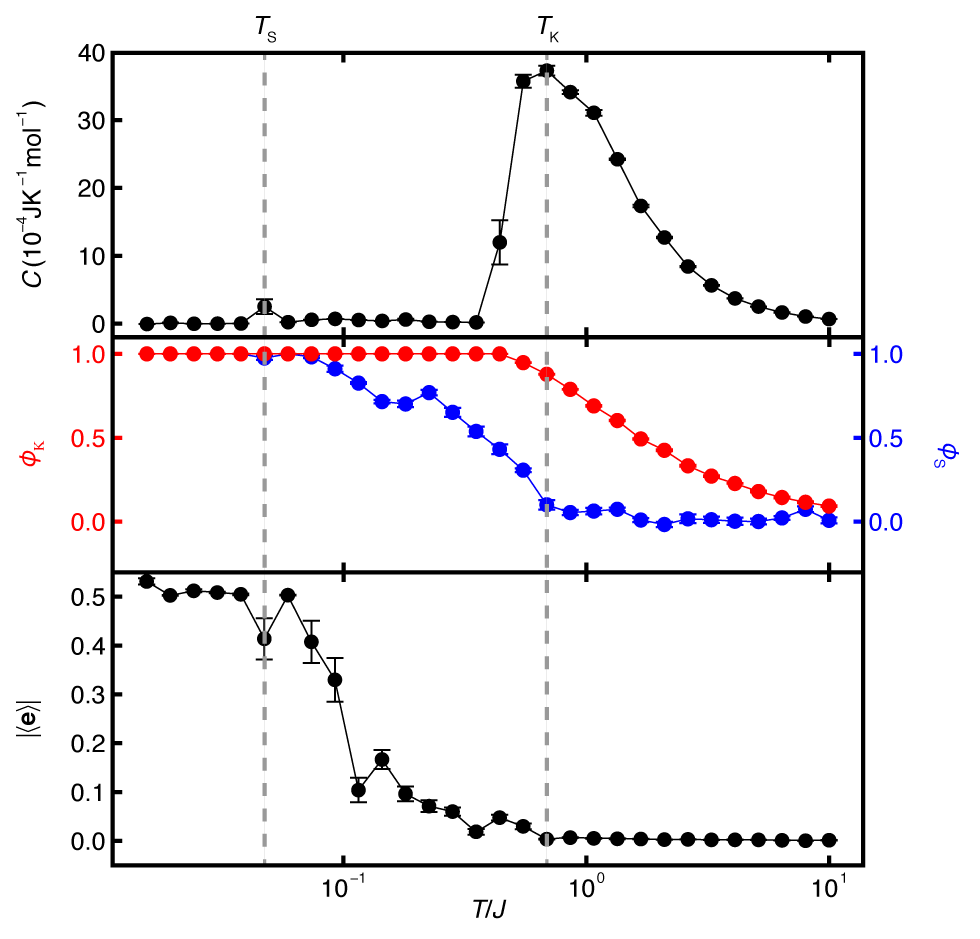}
\end{center}
\caption{\label{fig4} Temperature dependence of thermodynamic and local-order parameters for the intermediate case ($\theta=\pi/35$). The heat capacity (top) shows two temperature anomalies; the higher (at $T_{\textrm K}$) is associated with the onset of Kitaev-type local order, and the lower (at $T_{\textrm s}$) with the disappearance of local interactions of the type shown in Fig.~\ref{fig2}(b). The temperature dependence of the local order parameters is shown in the middle panel. The bottom panel shows the temperature dependence of the total polarisation magnitude, which saturates on cooling below $T_{\textrm s}$.}
\end{figure}

Monte Carlo simulations carried out for the intermediate case $J_2/J_1=9\%$ (\emph{i.e.}, $\theta=\pi/35$)---a value we will come to show is representative of the general case---indicate the presence of two specific heat anomalies [Fig.~\ref{fig4}]. Inspection of the temperature-dependence of the local order parameters $\phi_{\textrm K},\phi_{\textrm s}$ shows that the higher-temperature anomaly coincides with the onset of Kitaev order, and the lower-temperature anomaly with the onset of local order found in the steric interaction model described above [Fig.~\ref{fig4}]; hence we label the two transition temperatures as $T_{\textrm K}$ and $T_{\textrm s}$, respectively.

\begin{figure}[b]
\begin{center}
\includegraphics{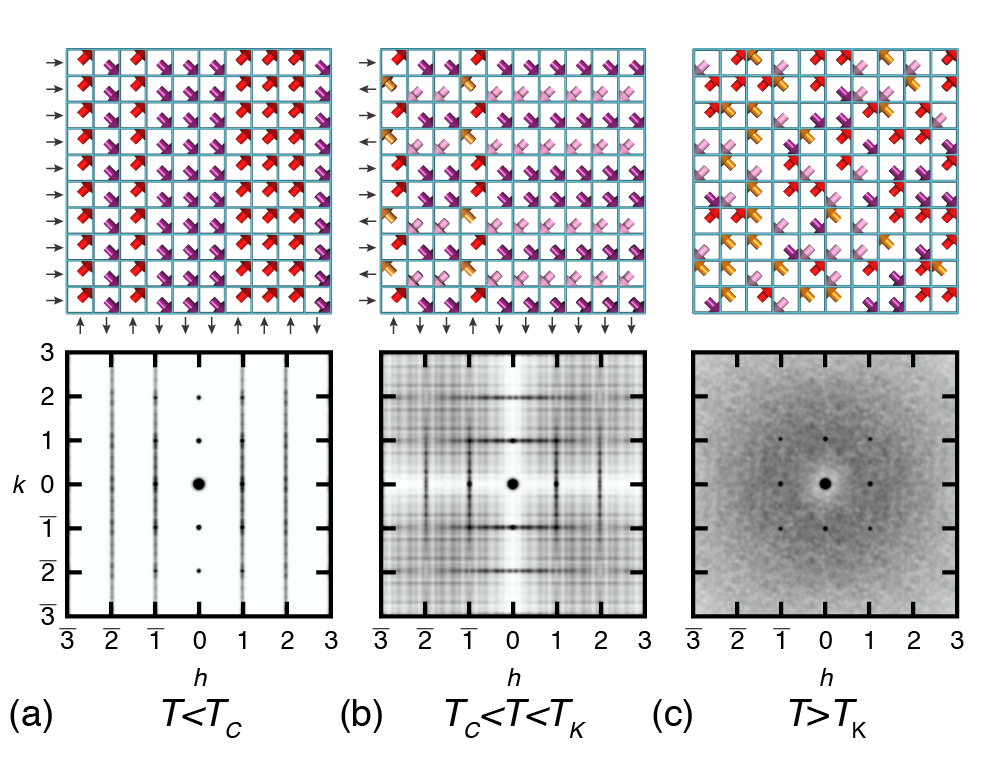}
\end{center}
\caption{\label{fig5} Representative displacement configurations and single-crystal scattering patterns for the three temperature regimes of the intermediate case ($\theta=\pi/35$) studied in Fig.~\ref{fig4}. (a) At low temperatures $T<T_{\textrm C}$ the system is polar; in the specific example shown here the bulk polarisation is oriented to the right-hand side. Note the persistence of configurational disorder in the vertical component of the polarisation for a given column of displacement vectors. This persistent disorder is reflected in the presence of streaks of diffuse scattering perpendicular to the $\mathbf a^\ast$ axis in reciprocal space. (b) The intermediate-temperature regime $T_{\textrm C}<T<T_{\textrm K}$ resembles the low-temperature behaviour of the pure-Kitaev model. The system is non-polar but strongly-correlated Kitaev-type local order gives rise to streaks of diffuse scattering perpendicular to both $\mathbf a^\ast$ and $\mathbf b^\ast$ axes in reciprocal space. (c) At high temperatures $T>T_{\textrm K}$ the system is paraelectric.}
\end{figure}

The key result of our study concerns the behaviour of the single-particle correlation function $\langle\mathbf e\rangle$ across these two transitions. This correlation function is a direct measure of bulk polarisation: $P=|\langle\mathbf e\rangle|$. As for the individual Kitaev and steric interaction models, we find no meaningful change in $P$ across $T_{\textrm K}$ [Fig.~\ref{fig4}]; hence this local-order transition is again hidden and is not associated with any long-range symmetry breaking. By contrast, the transition at $T_{\textrm s}$ involves the emergence of a non-zero bulk polarisation and so represents a Curie point with $T_{\textrm C}=T_{\textrm s}$. Consequently the phase behaviour for $0<\theta<\pi/4$ involves three regimes. At low temperatures $T<T_{\textrm C}$ the system is polar, with local order that reflects at once the ground states of both Kitaev and steric-interaction models. For intermediate temperatures $T_{\textrm C}<T<T_{\textrm K}$, the system is non-polar and resembles the low-temperature behaviour for $\theta=0$ (Kitaev order). Finally, at temperatures above $T_{\textrm K}$ local order is progressively lost and the system approaches the paraelectric limit. Representative configurations taken from these three regimes are shown in Fig.~\ref{fig5}, together with the corresponding diffraction patterns. As anticipated, the behaviour at $T_{\textrm K}$ closely resembles that observed in the pure-Kitaev limit ($\theta=0$).

The long-range symmetry-breaking process identified by the behaviour of the order parameter $P$ at $T_{\textrm C}$ is clearly evident in both real space and reciprocal space. We interpret this behaviour by contrasting against the low temperature behaviour of the pure-Kitaev model. In the Kitaev ground states each row or column has a common polarisation along the row/column axis; the system is disordered because these collective axial polarisations do not themselves order [Fig.~\ref{fig5}(a)]. This disordered state supports next-nearest neighbour displacements that act in direct opposition, which is evident in the value of $\phi_{\textrm s}<1$ for the Kitaev ground state [Fig.~\ref{fig3}]. For finite values of $\theta$, these opposing displacements are penalised by the Hamiltonian \eqref{Hamil}, and the polar ground state emerges as the mechanism by which the displacements are removed. As in the Kitaev ground state each row/column maintains a common polarisation, but collective polarisations are now ordered along one crystal axis [Fig.~\ref{fig5}(b)]. This order distinguishes one axis from the other and reduces the (average) plane group symmetry from $p4m$ to $pm$. From an average-structure perspective it is as if a polar distortion has condensed from a paraelectric parent phase. Yet, as we have seen, the state at temperatures just above $T_{\textrm C}$ supports strong correlations and is not truly paraelectric. Moreover, even the polar ground state remains disordered: the diffraction pattern contains continuous streaks of diffuse scattering parallel to the polar axis [Fig.~\ref{fig5}(b)].

\begin{figure}[t]
\begin{center}
\includegraphics{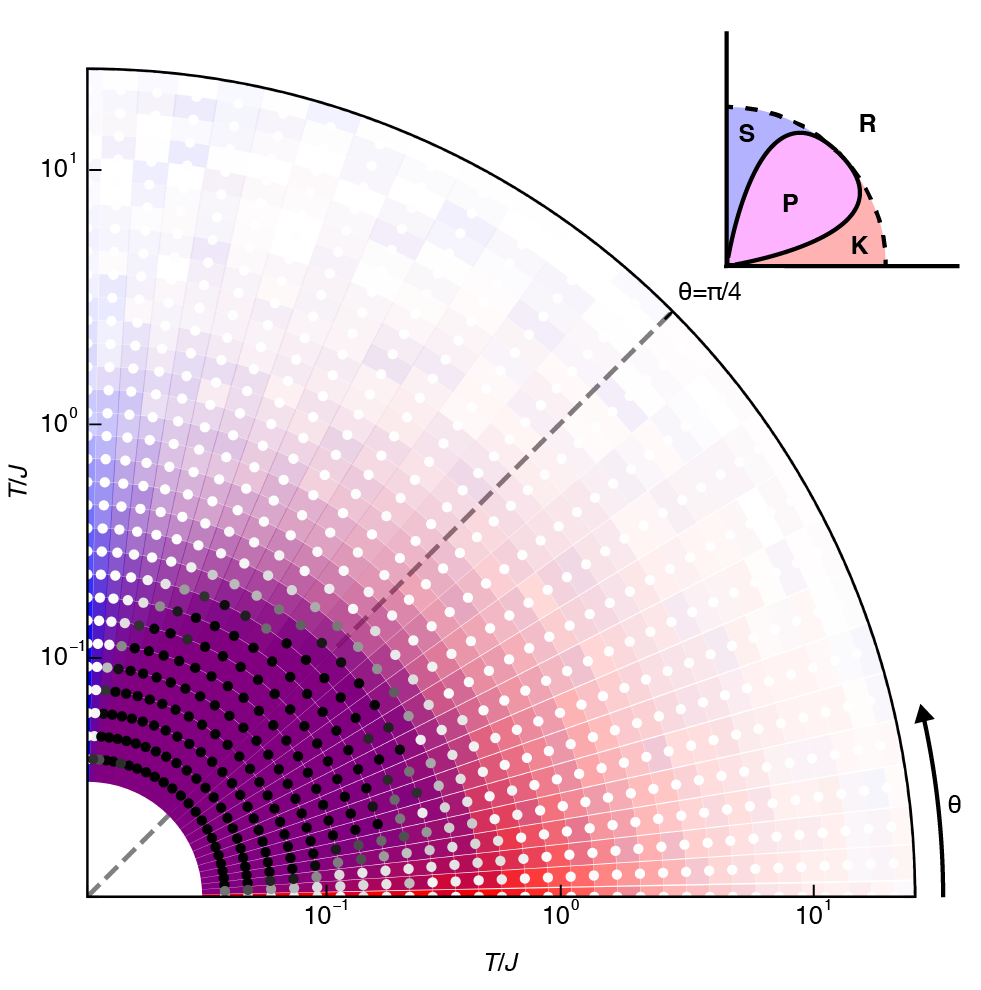}
\end{center}
\caption{\label{fig6} Phase diagram for the model described by Eq.~\eqref{Hamil}, given in terms of the polar coordinates $r=T/J$ and $\theta$ as defined in Eq.~\eqref{theta}. For each $(r,\theta)$ value the corresponding sector is coloured by the values of $\phi_{\textrm K}$ (red) and $\phi_{\textrm s}$ (blue). The corresponding polarisation magnitude is illustrated by the saturation of the small circle: $|\langle\mathbf e\rangle|=0$ (white) to $>0.5$ (black). Our key result is that polar states (black points) emerge from the combination of Kitaev- and steric-type local order (purple shading). Inset: ``local order phase diagram'', with Kitaev (K), steric (S), polar (P), and ``random'' (R) states indicated.}
\end{figure}

The phase behaviour driven by the Hamiltonian \eqref{Hamil} for other values of $\theta$ is conceptually very similar. When $\theta>\pi/4$ ($J_2>J_1$) the transition temperatures $T_{\textrm K},T_{\textrm s}$ swap their order, and the Curie temperature is now associated with $T_{\textrm K}$ rather than $T_{\textrm s}$. For $\theta=\pi/4$ the three transition temperatures coincide. For all values of $\theta$ between (but not equal to) the limits of 0 and $\pi/2$, the ground state is the polar disordered state identified above. A representation of the general phase behaviour is given in Fig.~\ref{fig6} as a radial function of $\theta$ and $T/J$. What is clear is that bulk polarisation emerges as the intersection of the locally-ordered low-temperature states of the Kitaev and steric interaction models.

For completeness we note that the use of periodic boundary conditions will necessarily introduce finite-size effects into our MC simulations. For example, the 1D order in the Kitaev state cannot be truly long-range at any finite temperature; instead any physical realisation must contain domain walls associated with Ising spin flips of this component. In terms of the polar phase that we find to emerge from combined Kitaev- and steric-type order, the implication is that ``bulk'' polarisation will persist for a given $\mathbf e$ only over a domain of finite (but arbitrarily large) size. The direction of $\mathbf e$ will vary from domain to domain, with domain size inversely proportional to temperature. This picture is of course entirely consistent with the presence of domains within the polar (ferroelectric) state in the absence of an external field.

\subsection{Extension to three dimensions}

Entirely analogous behaviour occurs in a three-dimensional variant of this same interaction model. The underlying lattice is now cubic rather than square, and the states representative of local polarisations parallel to one of the cube diagonals. Hence the $\mathbf e_{\mathbf r}\in\langle111\rangle$ can adopt one of eight vector values. The modified Hamiltonian is
\begin{eqnarray}\label{Hamil3d}
\mathcal H&=&-J_1\sum_{\mathbf r}\left(e_{\mathbf r}^xe_{\mathbf r+\mathbf a}^x+e_{\mathbf r}^ye_{\mathbf r+\mathbf b}^y+e_{\mathbf r}^ze_{\mathbf r+\mathbf c}^z\right)\nonumber\\& &+J_2\sum_{\mathbf r,\mathbf r^\prime}\delta\left[\mathbf e_{\mathbf r}^\parallel-\mathbf e_{\mathbf r^\prime}^\parallel-2(\mathbf r^\prime-\mathbf r)\right],\label{3dHamil}
\end{eqnarray}
with $\mathbf a,\mathbf b,\mathbf c$ the lattice vectors. The steric interaction sum is taken over pairs of next-nearest neighbours $\mathbf r,\mathbf r^\prime$, with the terms $\mathbf e_{\mathbf r}^\parallel,\mathbf e_{\mathbf r^\prime}^\parallel$ corresponding to the components of $\mathbf e_{\mathbf r},\mathbf e_{\mathbf r^\prime}$ along the axis connecting $\mathbf r$ and $\mathbf r^\prime$:
\begin{equation}
\mathbf e_{\mathbf r}^\parallel=\frac{\mathbf e_{\mathbf r}\cdot(\mathbf r^\prime-\mathbf r)}{|\mathbf r^\prime-\mathbf r|^2}(\mathbf r^\prime-\mathbf r).
\end{equation}
Again, the individual Kitaev and steric interaction terms each give rise to specific heat anomalies with no corresponding changes in macroscopic symmetry.

\begin{figure}
\begin{center}
\includegraphics{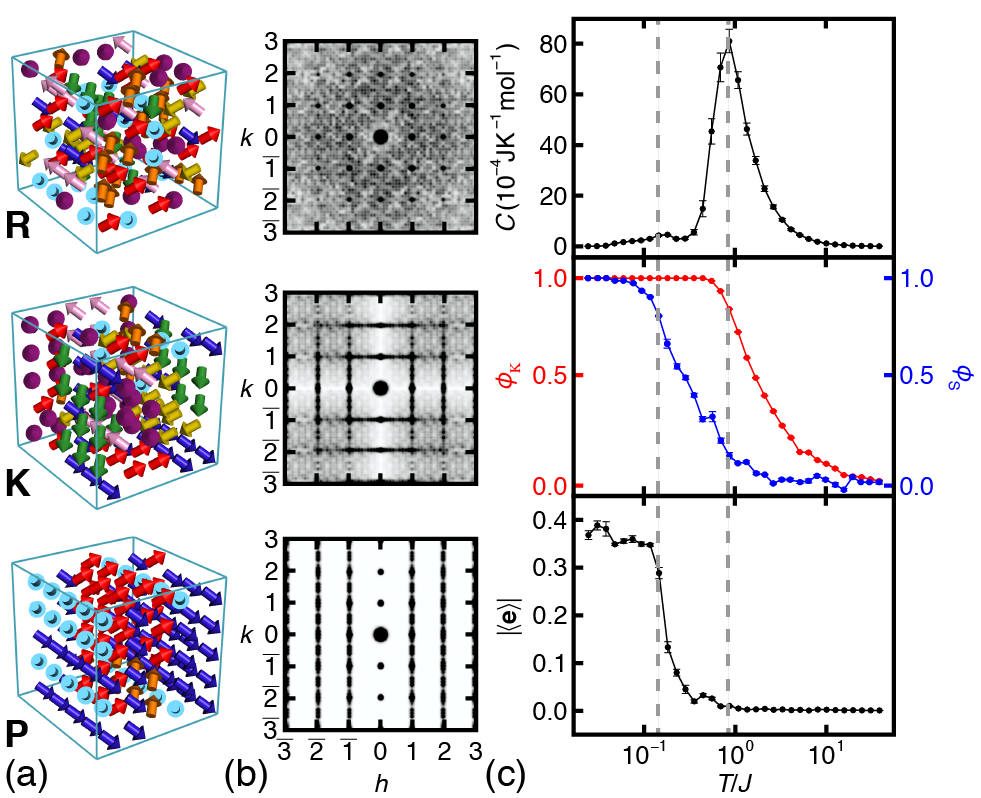}
\end{center}
\caption{\label{fig7} Phase behaviour of a representative example of the interaction model described by Eq.~\eqref{3dHamil}. (a) Displacement configurations within the high-temperature ``random'' (R) state, the intermediate-regime Kitaev (K) state, and the low-temperature polar (P) state. (b) The corresponding single-crystal scattering patterns show the emergence of increasingly structured diffuse scattering patterns as local order is increased. Note the reduced long-range symmetry of the P state. (c) Temperature dependence of the (top--bottom) specific heat, local order parameters, and bulk polarisation as determined by MC simulations.}
\end{figure}

As a representative example of the intermediate behaviour, we consider the case $J_1/J_2=25$. The temperature-dependence of the specific heat, shown in Fig.~\ref{fig7}, again reveals two phase transitions and hence identifies three phase fields. The nature of these phases is entirely analogous to that in the various two-dimensional models described above. The highest-temperature phase is paraelectric; the intermediate phase is non-polar and supports strong one-dimensional order of the Kitaev type. Both of these phases have $Pm\bar3m$ crystal symmetry. By contrast, the lower-temperature transition involves global symmetry breaking and leads to a ground-state polar phase with $P4mm$ symmetry. Again this state is heavily disordered, with collective axial degrees of freedom present along two of the three crystal axes. This transition is equivalent to the highest-temperature (cubic/tetragonal) transition reported in the MC study of BaTiO$_3$ in Ref.~\citenum{Senn_2016}.

\begin{figure}
\begin{center}
\includegraphics{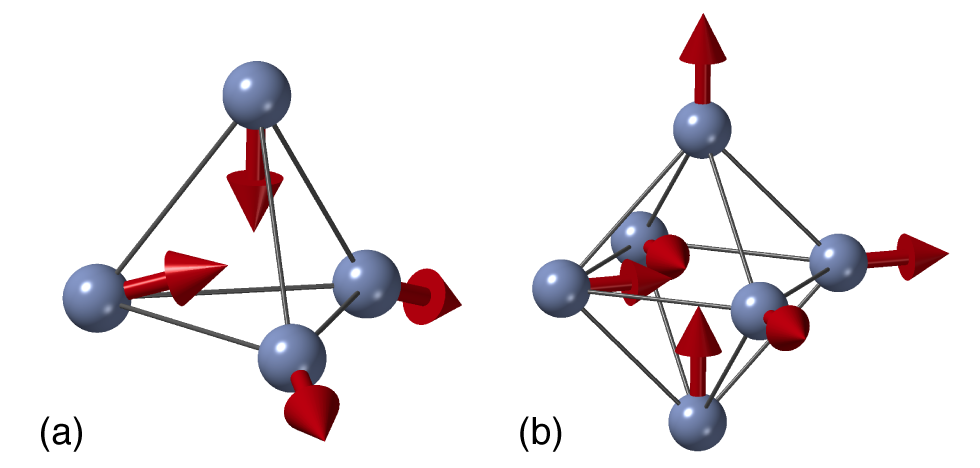}
\end{center}
\caption{\label{fig8} Ice-like states for simple polyhedra. (a) The ``2-in-2-out'' configuration of pyrochlore spin ices, and (b) the ``3-in-3-out'' configuration of the cubic Kitaev model.}
\end{figure}

We note for completeness that the Kitaev-like interaction term in Eq.~\eqref{Hamil3d} gives rise to a conceptually-interesting manifold of ``ice-like'' low-temperature states. Whereas conventional Ising spin ices are based on a topology of connected tetrahedra and obey the well-known ``2-in-2-out'' rules,\cite{Bramwell_1998} the simple-cubic ices that emerge from the Kitaev model are assembled from connected octahedra that obey a local ``3-in-3-out'' rule [Fig.~\ref{fig8}]. In cross section these configurations correspond to the square ice model;\cite{Ziman_1979,Keen_2015} moreover the states are related to the ``C$_3$F'' procrystalline phase described in Ref.~\citenum{Overy_2016} and the distribution of O and N atoms in magnetoresistive EuWO$_{1.5}$N$_{1.5}$.\cite{Yang_2010,Camp_2012} This is a model to which we expect to return in future studies.

\section{Concluding Remarks}

Returning to the context of inversion symmetry breaking in ferroelectrics we suggest that the Hamiltonian \eqref{Hamil3d} is likely relevant to the emergence of polarisation in well-studied systems such as BaTiO$_3$ and KNbO$_3$. Both single-crystal diffuse scattering\cite{Comes_1968,Comes_1970} and powder pair distribution function\cite{Senn_2016} measurements indicate that the ``paraelectric'' state of these systems just above $T_{\textrm C}$ is far from random and resembles instead the low-temperature behaviour of the Kitaev model. Hence we are in the $J_1>J_2$ regime of \eqref{Hamil3d}, which is sensible given the strong directional coupling between SOJT distortions expected for these systems [\emph{cf}.\ Fig.~\ref{fig2}]. We have shown that the instability that drives the $Pm\bar3m/P4mm$ transition--and hence the emergence of a spontaneous polarisation---in these two compounds need not involve ferroelectric coupling of the Ti/Nb displacements \emph{per se}, but may instead take the form of a local instability associated with steric or screened electrostatic interactions mediated by the A-site cations. Our goal here is not so much to explain the ferroelectric response of these systems, but rather to demonstrate that the combination of two judiciously-chosen local-order instabilities may in principle couple to a bulk polar distortion in a manner that is conceptually similar to the HIF mechanism.

One of the clear challenges that emerges from our study is the need for a much fuller understanding of the interplay between global and local symmetry breaking mechanisms. The conventional Landau description breaks down in attempting to describe local-order transitions (\emph{e.g.}\ at $T_{\textrm K}$ and $T_{\textrm s}$ in our study) but provides a very natural and appealing explanation of the global symmetry-breaking process that occurs at $T_{\textrm C}$. In other systems, local-structure transitions have been described in terms of Higgs transitions of an emergent gauge theory in an approach that is inherently non-Landau.\cite{Powell_2011} Consequently we suggest that the rationalisation of these two viewpoints is an appealing area for future theoretical study. In the context of materials design, the central question raised by our study is: what particular types of local ordering mechanisms might, when combined, necessarily couple to polarisation (or ferromagnetisation or ferroelasticity$\ldots$)? An ability to answer this question then informs the targeting of specific bonding motifs or components with specific local instabilities as an entirely new approach to functional materials discovery.

\begin{acknowledgments}
The authors gratefully acknowledge valuable discussions with N.~A.~Benedek (Cornell) and financial support from the E.R.C. (Grant 279705), the Royal Society (to M.S.S.), the Leverhulme Trust (Grant RPG-2015-292), and from Hertford College Oxford to E.H.W.
\end{acknowledgments}

\end{document}